\documentclass[pra,twocolumn,amsmath,amssymb,superscriptaddress,showpacs,floatfix]{revtex4-1}
\usepackage[latin9]{inputenc}
\usepackage{color}
\usepackage{amsmath}
\usepackage{amssymb}
\usepackage{esint}
\usepackage{dcolumn}
\usepackage{bm}
\usepackage{blindtext}
\usepackage{natbib}
\usepackage{graphicx}
\usepackage[hidelinks=true,pdfborder={0 0 0},colorlinks=true,
linkcolor=blue,urlcolor=blue,citecolor=blue]{hyperref}
\begin{document}
\title{Unconventional Bose-Einstein condensation in a system of two-species 
bosons in the $p$-orbital bands of a bipartite lattice}

\author{Jhih-Shih You}
\affiliation{Physics Department and Physics division, National Center 
for Theoretical Sciences, National Tsing-Hua University, Hsinchu 
30013, Taiwan}
\affiliation{Department of Physics, University of California, 
San Diego, CA92093}

\author{I-Kang Liu}
\affiliation{Department of Physics and Graduate Institute of Photonics, 
National Changhua University of Education, Changhua 50085, Taiwan}

\author{Daw-Wei Wang}
\affiliation{Physics Department and Physics division, National Center 
for Theoretical Sciences, National Tsing-Hua University, Hsinchu 30013, Taiwan}

\author{Shih-Chuan Gou}
\email{scgou@cc.ncue.edu.tw}
\affiliation{Department of Physics and Graduate Institute of Photonics, National
Changhua University of Education, Changhua 50085, Taiwan}

\author{Congjun Wu}
\email{wucj@physics.ucsd.edu}
\affiliation{Department of Physics, University of California, San Diego, CA92093}

\begin{abstract}
In the context of Gross-Pitaevskii theory, we investigate the unconventional 
Bose-Einstein condensations in the two-species mixture with $p$-wave symmetry 
in the second band of a bipartite optical lattice. 
A new imaginary-time propagation method is developed to numerically determine 
the $p$-orbital condensation. 
Different from the single-species case, the two-species boson mixture 
exhibits two non-equivalent complex condensates in the 
intraspecies-interaction-dominating regime, exhibiting the vortex-antivortex lattice configuration 
in the charge and spin channels, respectively.
When the interspecies interaction is tuned across the SU(2) invariant point, 
the system undergoes a quantum phase transition toward a checkerboard-like spin density wave state 
with a real-valued condensate wavefunction.
The influence of lattice asymmetry on the quantum phase transition is 
addressed. 
Finally, we present a phase-sensitive measurement scheme for experimentally 
detecting the UBEC in our model.
\end{abstract}
\pacs{03.75.Nt, 03.75.Lm, 05.30.Jp, 05.30.Rt}
\maketitle

\section{Introduction}

Unconventional condensate wavefunctions of paired fermions are identified by 
nontrivial representations of rotational symmetry, in contrast to the 
conventional counterpart with vanishing relative orbital angular momentum 
(OAM). 
Exploration of unconventional condensates dates back to the investigations 
of the $A$- and $B$-phases of the superfluid $^{3}$He
\cite{Anderson1961,Brinkman1974,Balian1963,Leggett1975} and later 
the spin-triplet pairing in Sr$_{2}$RuO$_{4}$
\cite{Maeno1994,Mackenzie2003,Nelson2004,Kidwingira2006}, which are 
characterized by the formation of Cooper pairs with OAM of $L=1$ and 
spin-triplet of $S=1$. 
High $T_{c}$ cuprates are another celebrated example whose pairing 
symmetry is $d_{x^{2}-y^{2}}$~\cite{wollman1993,Tsuei1994}.

Recently, considerable discoveries, both theoretical~\cite{Isacsson2005,
Liu2006,wu2006b,Kuklov2006,stojanovic2008,Collin2010,lixiaopeng2011,wu2009a,
cai2011,Martikainen2011,cai2012a,Xu2013,BoyangLiu2013,hebert2013,
XiaopengLi2014} and experimental~\cite{Sebby-Strabley2006,Mller2007,
wirth2011,olschlager2011,olschlager2012,olschlager2013,Kock2015,DongHu2015}, 
were reported on the single-boson condensation in the metastable high orbital 
bands of an optical lattice. 
The wavefunctions of this archetype of unconventional Bose-Einstein 
condensation (UBEC) are identified by the nontrivial representations 
of the lattice symmetry group, which oversteps the physical scenario set 
by ``no-node" theorem - an underlying principle of low-temperature 
physics stating that the many-body ground-state wavefunctions of Bose 
systems, including the superfluid, Mott-insulating and supersolid states, 
are necessarily positive-definite under general circumstances
\cite{Feynman1972,wu2009a}. 
In consequence, the wavefunctions of UBECs can be rendered complex-valued, 
and thus spontaneously break the time-reversal (TR) symmetry 
\cite{wu2009a}, which constitutes a remarkable feature of UBEC.
It is anticipated that UBECs can sustain exotic phenomena not seen in 
conventional BECs, such as the nontrivial ordering of OAM moment, BECs 
with nonzero momentum, half-quantum vortex and the spin texture of 
skyrmions~\cite{wu2009a}. 
It is also worth mentioning that the OAM moment formation still survives 
when system enters the Mott-insulating regime wherein the global 
U(1) phase coherence of superfluidity is no longer retained~\cite{cai2011}.

The experimental realization of single-species BECs in the second band, 
where the condensed atoms survive a long lifetime before tunnel to the 
nearly empty lowest band~\cite{wirth2011,olschlager2011,olschlager2012,
olschlager2013,Kock2015}, has marked an important progress towards the 
creation and manipulation of UBECs in ultracold atoms. 
Depending on the lattice asymmetry, the time-of-flight~(TOF) measurement 
revealed signatures of both real and complex condensates with $p$-wave 
symmetry and a large scale spatial coherence. 
The complex wavefunctions exhibit the configuration of a vortex-antivortex 
lattice with nodal points at vortex cores as theoretically predicted. 
More recently, a matter-wave interference technique was employed to provide
direct observations of the phase information of the condensate and to 
identify the spatial geometry of certain low energy excitations 
\cite{Kock2015}. 
The realization of UBECs in even higher bands  
was also reported~\cite{olschlager2011,olschlager2012}.

In this work, we present a theoretical study of the UBEC in a two-species 
boson mixture where both species are equally populated in 
the second band of a bipartite optical lattice~\cite{wirth2011}. 
Our study initiates the search of new types of UBECs enriched by coupling 
spin degrees of freedom with U(1) symmetry, TR symmetry and nontrivial 
representations of the lattice symmetry groups.  
To determine the wavefunction of the UBEC in the context of Gross-Pitaevskii 
equation~(GPE), we develop a numerical scheme which resorts to precluding 
the $s$-orbital components from the condensate wavefunction during the 
imaginary-time evolution of the full Hamiltonian. 
This scheme enables us to determine the phase diagram of UBEC in a wide 
range of parameters corresponding to the inter- and intraspecies interaction. 
We find that the emergent phases of UBEC  involve the $p_{x}\pm ip_{y}$ 
(complex-valued) and $p_{x}\pm p_{y}$ (real-valued) types of orbital order, 
which appear in different regimes of interaction that can be described  
as a consequence of the interplay between OAM and interaction energies, 
as will be discussed later.

This paper is organized as follows. 
The Sec.~\ref{sec:II}, we briefly account for the experimental setup of 
the bipartite two-dimensional lattice potential used in our model, including 
the symmetry analyses of the lattice configuration. 
The structure of the single-particle dispersion of the $p$-band is 
demonstrated. 
In Sec.~\ref{sec:III}, the numerical implementation of the modified 
imaginary-time propagation method is described, which, together with 
the Bloch wave approximation, enables to solve the GPEs in high bands. 
In Sec.~\ref{sec:IV}, we explore the properties of UBECs and phase 
transitions in the symmetric and asymmetric lattices. 
Finally,  a scheme for experimentally exploring the formation of UBECs 
in our model is addressed in Sec.~\ref{sec:V} and 
conclusions are made in Sec.~\ref{sec:VI}.

\section{The optical lattice and band spectrum}
\label{sec:II}

We consider the two-species BEC in the first excited orbital band of the 
bipartite optical lattice employed in the experiments~\cite{wirth2011}, 
where the unit cell consists of two sites with different potential depths. 
The optical potential $V(x,y)$ is described by
\begin{equation}
\begin{array}{ll}
{\displaystyle V(\mathbf{r})=-\frac{V_{0}}{4}\left|\eta
\left[\left(\mathbf{e}_{z}\cos\alpha+\mathbf{e}_{z}\sin\alpha\right)e^{ik_{l}x}
\right.\right.}\\
{\displaystyle \qquad\qquad\left.\left.+\mathbf{e}_{z}\epsilon 
e^{-ik_{l}x}\right]+\mathbf{e}_{z}e^{i\theta}\left(e^{ik_{l}y}
+\epsilon e^{-ik_{l}y}\right)\right|^{2},}
\end{array}\label{eq:Hemmerich_potential}
\end{equation}
where the unit vectors $\mathbf{e}_{z}$ and $\mathbf{e}_{y}$ constitutes 
the basis of the light polarization; 
$V_{0}$ is determined by the laser power; 
$k_{l}=2\pi/a_{0}$ is the laser wavevector; 
$\alpha$ is the polarization angle with respect to $z$-direction; 
$\epsilon$ is the reflection loss; 
the intensity and phase differences between laser beams along 
the $x$- and $y$-directions are described by $\eta$ and $\theta$, 
respectively. 
The symmetry analysis of the lattice configuration and the subsequent 
band structure calculations have already been presented 
in Ref.~[\onlinecite{cai2011}]. 
Below we recap this analysis in detail to make the paper self-contained.

\begin{figure}[bhtp]
\begin{center}
\includegraphics[width=1\linewidth]{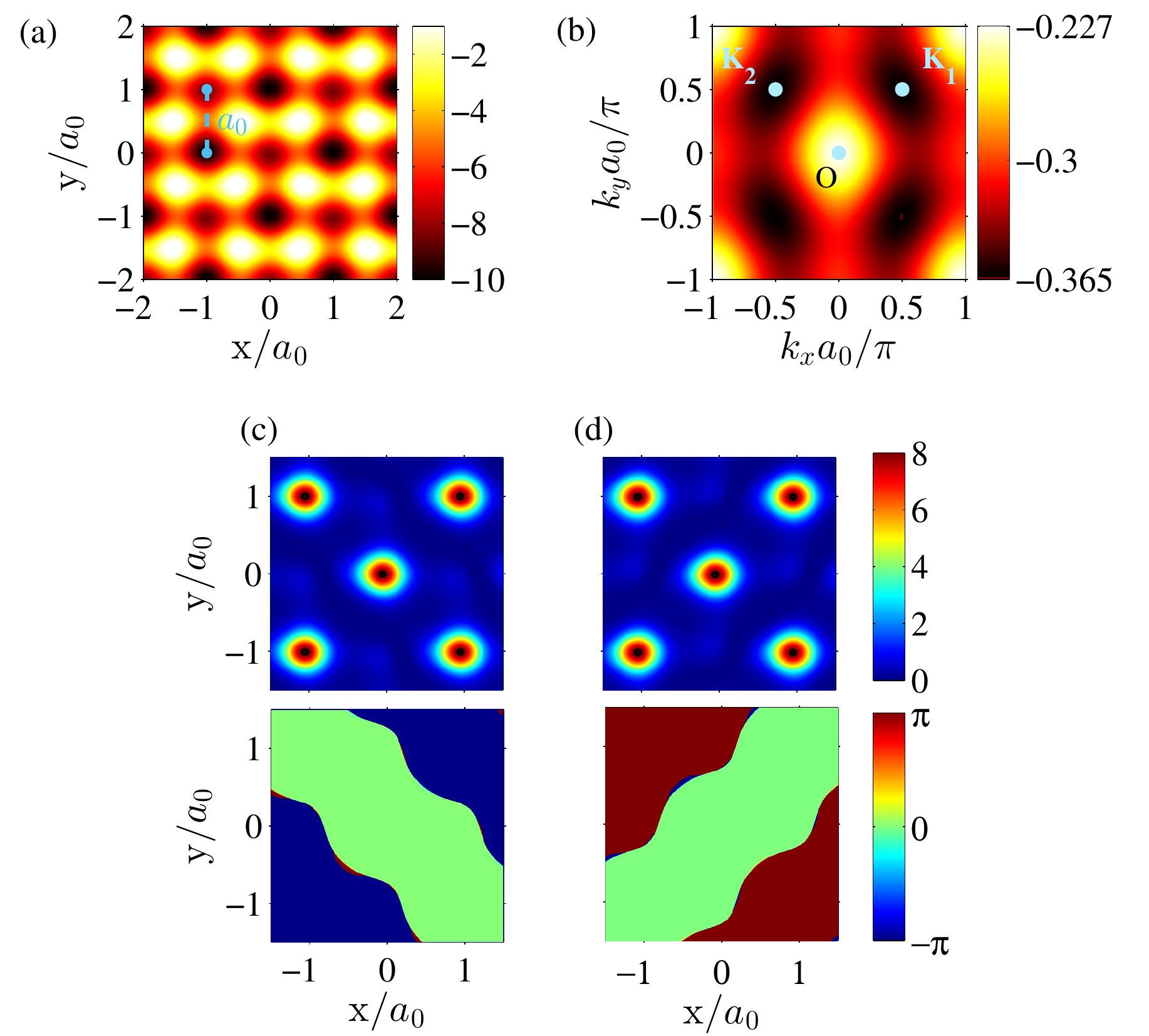}
\caption{(a) The double-well optical lattice with the
experiment parameters $V_{0}=6.2E_{r}$, $\eta=0.95$,
$\theta=95.4^\circ$, $\epsilon=0.81$, and $\alpha=\pi/5$.
The white dashed line illustrates the half wavelength of
laser $a_0$  and $\sqrt{2}a_0$ is the lattice constant.
The $A$ and $B$ sublattice sites are denoted in (a).
(b) The energy spectra of the second band, whose
energy minima are located at $\mathbf{K}_{1}=({\pi}/{2a_0}, {\pi}/{2a_0})$
and $\mathbf{K}_{2}=(-{\pi}/{2 a_0},{\pi}/{2 a_0})$. 
(c) and (d) are the density profiles (upper panel) and phase profiles 
(bottom panel) for non-interacting gas for $\mathbf{K}_{1}$ and 
$\mathbf{K}_{2}$, respectively.
}
\label{potential}
\end{center}
\end{figure}

For the ideal case with $\eta=1$, $\epsilon=1$, and $\alpha=0$, the 
lattice potential is simplified as
\begin{equation*}\begin{array}{rl}
V(\mathbf{r}) = & -V_{0}\Big(\cos^{2}k_{l}x+\cos^{2}k_{l}y\\
&\qquad\quad+2\cos k_{l}x\cos k_{l}y\cos\theta\Big),
\end{array}\end{equation*}
which possesses the tetragonal symmetry. Since $\theta$ controls 
the relative depth of the double-well inside the unit cell, 
tuning $\theta$ away from $90^\circ$ results in the bipartite lattice. 
When $\eta<1$ and $\epsilon=1$, the lattice potential becomes
\begin{equation*}\begin{array}{rl}
V(\mathbf{r}) =&-V_{0}\Big(\eta^{2}\cos^{2}k_{l}x+\cos^{2}k_{l}y\\
&\qquad\quad+ 2\eta\cos k_{l}x\cos k_{l}y\cos\theta\Big),
\end{array}\end{equation*}
which still possesses the reflection symmetries with respect to both 
$x$ and $y$-axes, but the point group symmetry is reduced to the 
orthorhombic one. 
For the realistic case with $\eta<1$ and $\epsilon<1$, the orthorhombic 
symmetry is broken such that in general no special point group symmetry 
survives. 
Nevertheless, the lattice asymmetry can be partially restored at 
$\alpha_0=\cos^{-1} \epsilon$, where the lattice potential becomes
\begin{equation*}\begin{array}{rl}
V(\mathbf{r})=&\displaystyle-\frac{V_{0}}{4}\Big\{ (1+\eta^{2})(1+\epsilon^{2})
+2\epsilon^{2}\eta^{2}\cos2k_{l}x\\
&\displaystyle+2\epsilon^{2}\eta^{2}\cos2k_{l}y\\
&\displaystyle+4\epsilon\eta\cos2k_{l}x
\Big[\epsilon\cos(k_{l}y-\theta)+\cos(k_{l}y+\theta)\Big]\Big\}
\end{array}\end{equation*}
and the reflection symmetry with respect to the $y$-axis is retrieved.
Therefore we call the case of $\alpha$ with $\alpha\neq\alpha_{0}$ as 
``asymmetric'' and that of $\alpha=\alpha_{0}$ as ``symmetric'', respectively. 
The lattice structure with the experimental parameters $V_{0}=6.2E_{r}$, 
$\eta=0.95$, $\theta=95.4^\circ$, $\epsilon=0.81$, and $\alpha=\pi/5$ 
is shown in Fig.~\ref{potential}~($a$).

The Bloch-wave band structure of the Hamiltonian $H_0=-\hbar^{2}\nabla^{2}/2M+V(\mathbf{r})$
can be calculated based on the plane-waves basis. The reciprocal lattice vectors are defined as $\mathbf{G}_{m,n}=m\mathbf{b}_{1}+n\mathbf{b}_{2}$ with $\mathbf{b}_{1,2}=(\pm\pi/a_0,\pi/a_0)$ where $a_0$ is a half wavelength of the laser. The diagonal matrix elements are
\begin{equation}\begin{array}{llll}
\displaystyle\langle\mathbf{k}+\mathbf{G}_{m,n}|H_{0}|\mathbf{k}+\mathbf{G}_{m,n}\rangle=E_{r}\\
\;\displaystyle\times\left\{\left[\frac{a_{0}k_{x}}{\pi}+(m-n)\right]^{2}+\left[\frac{a_{0}k_{y}}{\pi}+(m+n)\right]^{2}\right\}
\end{array}\end{equation}
where $\mathbf{k}$ is the quasi-momentum in the first Brillouin zone, and the off-diagonal matrix elements are
\begin{equation}\begin{array}{ll}
\displaystyle\langle\mathbf{k}|V|\mathbf{k}+\mathbf{G}_{\mp1,0}
\rangle\;\;=-\frac{V_{0}}{4}\epsilon\eta(e^{\pm i\theta}
+\cos\alpha e^{\mp i\theta}),\\
\\
\displaystyle\langle\mathbf{k}|V|\mathbf{k}
+\mathbf{G}_{0,\pm1}\rangle\;\;
=-\frac{V_{0}}{4}\eta(\epsilon^{2}e^{\pm i\theta}+\cos\alpha e^{\mp i\theta}),\\
\\
\displaystyle\langle\mathbf{k}|V|\mathbf{k}+\mathbf{G}_{\mp1,\pm1}\rangle=-\frac{V_{0}}{4}\eta^{2}\epsilon\cos\alpha,\\
\\
\displaystyle\langle\mathbf{k}|V|\mathbf{k}+\mathbf{G}_{\mp1,\mp1}\rangle=-\frac{V_{0}}{4}\epsilon.
\end{array}
\label{potential_elements}
\end{equation}
The energy spectrum of the second band of the optical lattice, 
Eq.~(\ref{eq:Hemmerich_potential}),  is shown in Fig.~\ref{potential}~($b$). 
Several observations are in order. 
Firstly, The energy minima are located at 
$\mathbf{K}_{1,2}\equiv\mathbf{b}_{1,2}/2$ with the corresponding 
wavefunctions $\psi_{\mathbf{K}_{1}}$ and $\psi_{\mathbf{K}_{2}}$.  
For the symmetric lattice, $\psi_{\mathbf{K}_{1}}$ and $\psi_{\mathbf{K}_{2}}$ 
are degenerate due to reflection symmetry, while for the asymmetric lattice, 
the degeneracy is lifted. 
Secondly, there are four points in the Brillouin zone (BZ), namely, 
the zero center $O$, the high symmetry point $X$, $(\pi/a_{0},\pi/a_0)$, 
and $\mathbf{K}_{1,2}$, which are TR invariant because their
opposite wavevectors are equivalent to themselves up to reciprocal 
lattice vectors. As a result, their Bloch wavefunctions are real, in other words, they are standing waves instead of propagating waves. Thirdly, the hybridized nature of $\psi_{\mathbf{K}_{1}}$ and $\psi_{\mathbf{K}_{2}}$ is also manifest in real-space: their wavefunctions are mostly in the superposition of the local $s$-orbital of the shallow well and the $p$-orbital of the deep well, which possess nodal lines passing through the centers of the deeper wells as shown in Fig. \ref{potential}~($c$) and ($d$).

\section{The modified imaginary-time propagation method}
\label{sec:III}
In current experiments \cite{wirth2011}, the correlation effects are 
relatively weak due to the shallow optical potential depth, and thus the 
two-species UBEC can be well described by the coupled GPE as
\begin{equation}
E\Psi_{\beta}(\mathbf{r})=\left[H_{\beta}^{0}+\sum_{\alpha=A,B}
\tilde{g}_{\beta\alpha}|\Psi_{\alpha}(\mathbf{r})|^{2}\right]\Psi_{\beta}(\mathbf{r})
\label{eq:CGPE}\end{equation}
where $H_{\beta}^{0}=(-\hbar^{2}\nabla^{2})/{2M_{\beta}}+V(\mathbf{r})$ is 
the one-particle Hamiltonian and the wavefunction $\Psi_{\beta}$ is normalized 
to the area of one unit cell, $\int'd^{2}r|\Psi_{\beta}(\mathbf{r})|^{2}
=\Omega=2a_{0}^{2}$; $\tilde{g}_{\alpha\beta}=g_{\alpha\beta}n_{\beta}$ 
with $n_{\beta}$ the particle number per unit cell and $g_{\alpha\beta}$ 
the interaction strength between $\alpha$ and $\beta$ species. 

Furthermore, in terms of $\Psi_{A}$ and $\Psi_{B}$, the real-space spin 
density distribution is defined as $\mathbf{S}(\mathbf{r})=(1/2)
\Psi^{\dagger}(\mathbf{r})\hat{\sigma}\Psi(\mathbf{r})$ 
where $\Psi\equiv(\Psi_{A},\Psi_{B})^{T}$ and $\hat{\sigma}$ denotes 
the Pauli matrices in vector form. 
Explicitly, the Cartesian components of the spin density are 
related to $\Psi_{A}$ and $\Psi_{B}$ by 
$S_{x}+iS_{y}=\sqrt{2}\hbar\Psi_{A}^{\ast}\Psi_{B}$ and 
$S_{z}=\hbar\left(|\Psi_{A}|^{2}-|\Psi_{B}|^{2}\right)$. 
Obviously, the orientation of spin in $xy$ plane  depends only on 
the global phases of $\Psi_{A}$ and $\Psi_{B}$.

In solving Eq.~(\ref{eq:CGPE}), we assume $\tilde{g}_{AA}=\tilde{g}_{BB}$, 
$\tilde{g}_{AB}=\tilde{g}_{BA}$,  $M_{A}=M_{B}=M$~\cite{myatt1997,Hall1998}, 
and $n_{A}=n_{B}$.
Since the band minima are located at $\mathbf{K}_{1,2}$, we expand the 
two-species condensate wavefunction in terms of $\psi_{\mathbf{K}_{1}}$ 
and $\psi_{\mathbf{K}_{2}}$,
\begin{equation}
\left(\begin{array}{l}
\Psi_{A}(\mathbf{r})\\
\Psi_{B}(\mathbf{r})
\end{array}\right)=\left(\begin{array}{c}
\cos\delta_{A}\psi_{\mathbf{K}_{1}}(\mathbf{r})+e^{i\phi_{A}}
\sin\delta_{A}\psi_{\mathbf{K}_{2}}(\mathbf{r})\\
\cos\delta_{B}\psi_{\mathbf{K}_{1}}(\mathbf{r})
+e^{i\phi_{B}}\sin\delta_{B}\psi_{\mathbf{K}_{2}}(\mathbf{r})
\end{array}\right).
\label{eq:bloch_expansion}\end{equation}
In general, $\psi_{\mathbf{K}_{1}}(\mathbf{r})$ and $\psi_{\mathbf{K}_{2}}
(\mathbf{r})$ are determined by the renormalized lattice potential, 
and are thus different from those based on the free band Hamiltonian 
$H_{0}$~\cite{species_dependent}.
Because the particle number of each species is conserved separately, 
the formation of two-species BEC spontaneously breaks the U(1)$\times$U(1) 
symmetry, leaving the freedom of choosing the condensate wavefunction 
by individually fixing the phase factor of  $\psi_{\mathbf{K}_{1}}(\mathbf{r})$ 
in each species of Eq.~(\ref{eq:bloch_expansion}).

The theoretical model in the single-species UBEC based on the GP description 
has been investigated with a self-consistent approach~\cite{cai2011,Xu2013}. 
For the two-species case, the structure of competing orders is even richer 
than that of the single-species case. In the enlarged phase space, the 
orbital states can entwine with spin degrees of freedom. 
We introduce a modified imaginary-time propagation method to solve the 
two-species UBEC, which liberates us from the restriction of certain 
types of solutions and can be generalized to other higher orbital bands 
as well. 
Since the ordinary imaginary-time propagation method only applies to yield 
the ground-state condensate, in order to reach the UBEC in the second band, 
the new method is devised to constantly project the lower orbital components 
out of the evolving (in imaginary time) condensate wavefunction, forcing 
the initial wavefunction evolve to the stationary solution in the target 
orbital. 
We have examined this method for one- and two-dimensional harmonic 
oscillators, and the resultant wavefunctions not only converge to the 
exact solutions, but also yield the correct degeneracy of high 
energy levels.

The implementation of our imaginary-time propagation algorithm is summarized 
as follows. We start by initializing a trial condensate wavefunction in the 
form of Eq.~(\ref{eq:bloch_expansion}) with $\psi_{\mathbf{K}_{1,2}}$ determined 
by $V(\mathbf{r})$ of the empty lattice. 
After the propagation of one time step, we arrive at a new set of $\Psi_{A}$ 
and $\Psi_{B}$ which are then employed to generate the renormalized lattice 
potential $V_{eff,\alpha} (\mathbf{r})=V(\mathbf{r})
+\sum_{\beta}\tilde{g}_{\beta\alpha}|\Psi_{\alpha}(\mathbf{r})|^{2}$. 
Then we solve the $s$-orbital states $|\varphi_{\mathbf{k}}\rangle_{\alpha}$
at  $\mathbf{k}=\mathbf{K}_{1}$ and $\mathbf{K}_{2}$ based on $V_{eff,\alpha}$, 
and construct the projection operator  
\begin{equation}
\hat{P}=1-\sum_{\alpha=A, B}\sum_{\mathbf{k}=\mathbf{K}_{1},\mathbf{K}_{2}}|
\varphi_{\mathbf{k}}\rangle_{\alpha\;\alpha}\langle \varphi_{\mathbf{k}}|. 
\end{equation}
After projecting out the $s$-orbital component by applying $\hat{P}$ 
to $\boldsymbol{\Psi}$, we proceed to the next step of imaginary-time 
evolution. 
The above process is repeated until the convergence is achieved and 
$\hat{P}$ is updated in each step. 
To assure the reliability of this method, we choose several different 
initial trial wavefunctions and add small complex random noises to 
break any specific symmetry which could lock the solution. 
Every simulation was implemented with a sufficiently long time to ensure 
that the energy converges. 
We have successfully reproduced the one-species UBEC solutions in the 
second band, and confirmed the results consistent with the previous 
works \cite{cai2011,Xu2013}. 
The interaction strengths are much smaller than the energy difference 
between the $s$ and $p$-orbital bands in our simulations, and 
thus the band mixing effect is negligible.

\section{Main results}\label{sec:IV}

\subsection{The symmetric lattice}\label{sec:IVa}
We first consider the symmetric lattice and the competition between intra- 
and interspecies interactions which determines the condensate wavefunctions. 
Defining $\gamma=\tilde{g}_{AB}/\tilde{g}_{AA}$, we start with 
an SU(2) symmetry breaking case in the regime of $\gamma<1$.
When $\gamma=0$, the system simply reduces to two decoupled single-species 
problems and each of them is in the complex condensate exhibiting 
nodal points rather than nodal lines. 
Accordingly, there are two $p$-orbital condensations characterized 
by substituting the following phase angles into 
Eq.(\ref{eq:bloch_expansion}): 
(I) $\phi_{A}=\phi_{B}=\pm\frac{\pi}{2}$, $\delta_{A}=\delta_{B}=\frac{\pi}{4}$,
\begin{equation}
\left(\begin{array}{c}
\Psi_{A}(\mathbf{r})\\
\Psi_{B}(\mathbf{r})
\end{array}\right)=\frac{1}{\sqrt{2}}\left(\begin{array}{c}
\displaystyle\psi_{\mathbf{K}_{1}}(\mathbf{r})+i\psi_{\mathbf{K}_{2}}(\mathbf{r})\\
\displaystyle\psi_{\mathbf{K}_{1}}(\mathbf{r})+i\psi_{\mathbf{K}_{2}}(\mathbf{r})
\end{array}\right),
\label{eq:solution3}\end{equation}
and (II) $\phi_{A}=-\phi_{B}=\pm\frac{\pi}{2}$, $\delta_{A}=\delta_{B}=
\frac{\pi}{4},$
  \begin{equation}
\left(\begin{array}{c}
\Psi_{A}(\mathbf{r})\\
\Psi_{B}(\mathbf{r})
\end{array}\right)=\frac{1}{\sqrt{2}}\left(\begin{array}{c}
\displaystyle\psi_{\mathbf{K}_{1}}(\mathbf{r})+i\psi_{\mathbf{K}_{2}}(\mathbf{r})\\
\displaystyle\psi_{\mathbf{K}_{1}}(\mathbf{r})-i\psi_{\mathbf{K}_{2}}(\mathbf{r})
\end{array}\right).
\label{eq:solution3}\end{equation}
When $0<\gamma<1$,  the corresponding $p$-orbital solutions take the forms 
of state (I) and (II) as well.

\begin{figure}[hbtp!]
\includegraphics[width=1\linewidth]{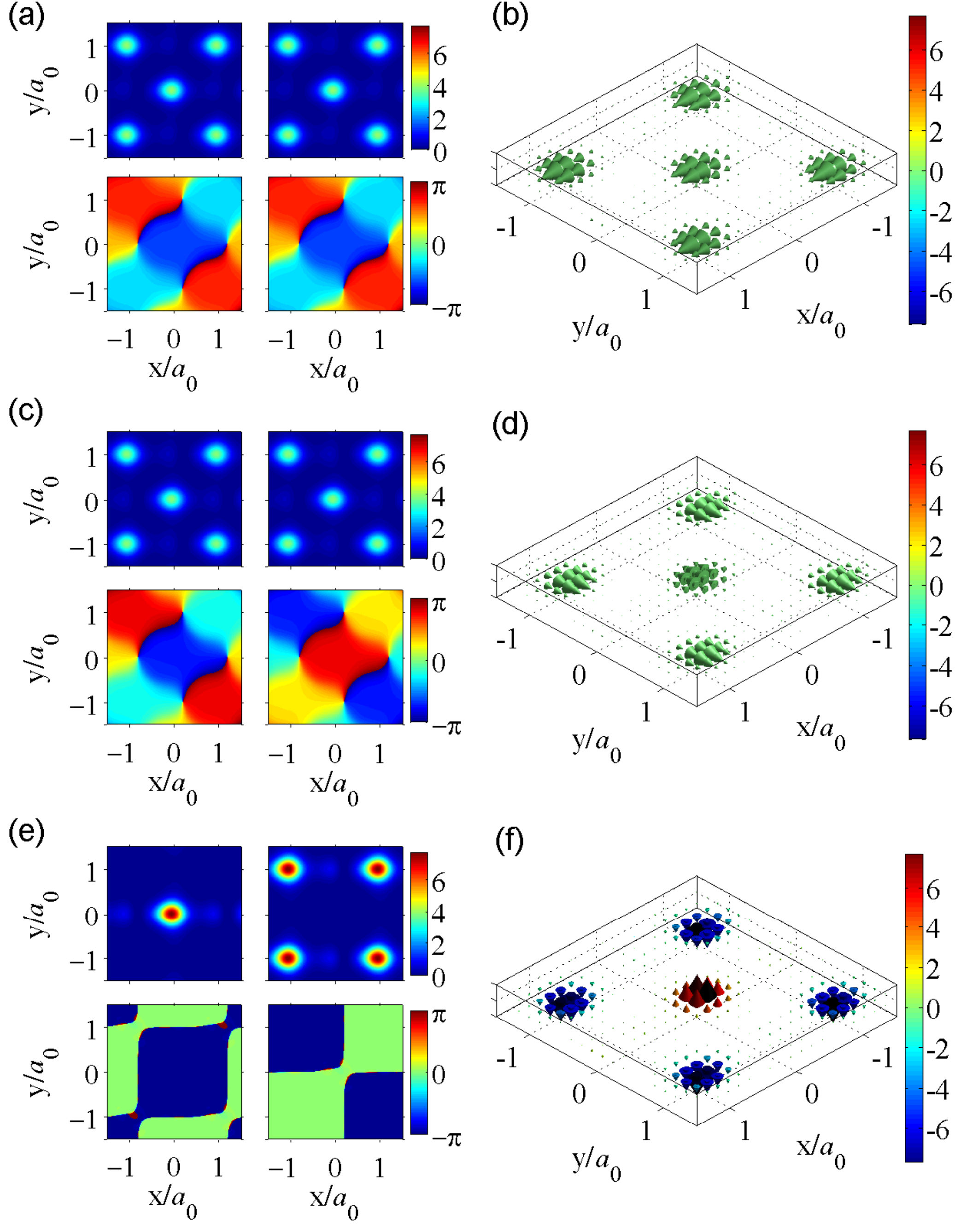}
\centering
\caption{Spatial distribusions of the density, phase, and spin texture of the condensate
wavefunctions corresponding to various states in the symmetric lattice are
showcased in different groups of subplots: state~(I)~[(a) and~(b)], state~(II)~[(c)
and (d)] and {\it{checkerboard}} state [(e) and (f)].
 The intra-species interactions are $\tilde{g}_{AA}=\tilde{g}_{BB}=0.025 E_r$  
with $E_{r}=\hbar^{2}k_{l}^{2}/2M$; the interspecies ones are $\tilde{g}_{AB}=0.25 \tilde{g}_{AA}$ for ($a$) and ($b$), and $\tilde{g}_{AB}=1.1\tilde{g}_{AA}$ for ($c$) and ($d$), respectively. In ($a$), ($c$) and ($e$), the density and phase distributions are shown in the upper and lower panels for each species, respectively. The spin texture configurations are shown in ($b$), ($d$) and ($f$), respectively, with arrows indicating the orientation of spins and color bars representing the values of $S_{z}$. The parameters used are $V_{0}=6.2E_{r}$, $\eta=0.95,$ $\theta=95.4^\circ,$ $\epsilon=0.81$, $\alpha=\alpha_{0}=\cos^{-1}\epsilon\approx35.9^{\circ}$. }
\label{fig:density_phase_spin}
\end{figure}

States (I) and (II) possess different symmetry structures, as illustrated in Fig.~\ref{fig:density_phase_spin} ($a$) to ($d$). 
Species $A$ and $B$ can be interpreted as a Kramers doublet, and a commonly 
used Kramers-type TR transformation is defined as $\hat{T}=i\hat{\sigma}_{y} \hat{C}$ 
where $\hat{C}$ is complex conjugation operation and $\hat{\sigma}_{y}$ is the Pauli matrix.
$\hat{T}$ keeps particle number and spin-current invariant but flips the sign of 
spin and charge current, and it satisfies $\hat{T}^2=-1$.
For state (I), its axial OAM moments of two species are parallel exhibiting
a vortex-antivortex lattice configuration, the condensate spin is polarized 
along the $x$-direction, which obviously breaks Kramers TR symmetry.
As for state (II), its axial OAM moments are antiparallel to each other
exhibiting a spin-current vortex-antivortex lattice configuration.
Although spin current is invariant under Kramers TR transformation,
the spin density exhibits the in-plane spin texture with the winding
number $\pm 2$ around each vortex core, which also breaks the Kramers
TR symmetry.
Nevertheless, state (II) is invariant by the anti-linear transformation
$\hat{T}^\prime=\hat{\sigma}_x \hat{C}$, which is equivalent to a combination of the TR 
transformation followed by a rotation around the $z$-axis at $\pi$.
Since $T^{\prime,2}=1$, it is no longer a Kramer transformation, which 
maintains the $xy$-components of spin invariant but flips the 
$z$-component of spin.

States (I) and (II) give rise to the same particle density and kinetic 
energy distributions for both species, and thus their energy are degenerate
at the mean-field GPE level.
Nevertheless, since they are not directly connected by symmetry, 
this degeneracy is accidental and only valid at the GPE level. 
The system symmetry allows a current-current interaction between two species, 
which is absent in the bare Hamiltonian, but could be effectively generated 
through quantum fluctuations for low energy physics in the sense of 
renormalization group. 
Since the current density distributions of two species are the same 
in one solution but are opposite in the other. 
This emergent interaction would lift this accidental degeneracy. 
However, this is a high order effect beyond the GPE level, which is 
certainly an interesting subject for future investigations.

The spatial distributions of the population and phase of both condensate 
species, together with the corresponding spin texture are shown in 
Fig.~\ref{fig:density_phase_spin} ($a$) and ($b$), respectively.
The particle density mainly distributes in the shallow sites which is the 
nodeless region corresponding to the $s$-orbital, while the density in the 
deep sites at which the nodal points are located corresponding to 
the $p_{x(y)}$-orbitals.
Each species exhibits a vortex-antivortex lattice structure: 
The vortex cores are located at the deep sites, and the nodeless region 
exhibits the quadripartite sublattice structure featuring the cyclic phase 
factors $\exp\left(i\pi n/2\right)$ for $n\in\{1,2,3,4\}$ in the 
shallow sites. 
For state (I), both species exhibit the same vorticity distribution and 
thus the spin density orientation lies along the $x$-direction according 
to the phase convention of Eq.~(\ref{eq:bloch_expansion}). 
There is no preferential direction of spin orientation in $xy$ plane 
due to  U(1) symmetry generated by the total $z$-component spin. 
For state (II), the two species exhibit opposite vorticities, and 
the configuration is a spin-vortex-antivortex lattice. 
In both cases, the vorticity or the spin vorticity patterns exhibit 
a double period of the lattice potential.

With $\gamma=1$, the sum of interaction energies is rendered an 
SU(2)-invariant form such that the wavefunctions of UBEC become 
highly degenerate. 
At this point, the states (I) and (II) persist as expected.
Because of the SU(2) invariance, we can further apply the global SU(2) 
rotations to states (I) and (II)~\cite{su2rotation}. 
For state (I), the constraint of maintaining $n_{A}=n_{B}$ does not 
allow new states under the form of Eq.~(\ref{eq:bloch_expansion}). 
For state (II), any SU(2) rotation still maintains $n_{A}=n_{B}$. 
For example, after a rotation of $-\pi/2$ around the $y$-axis, 
we arrive at $(\Psi_{A},\Psi_{B})=(\psi_{\mathbf{K}_{1}},-i\psi_{\mathbf{K}_{2}})$, 
and a subsequent $\pi/2$-rotation around the $x$-axis yields
\begin{equation}
\left(\begin{array}{c}
\Psi_{A}(\mathbf{r})\\
\Psi_{B}(\mathbf{r})
\end{array}\right)=\frac{1}{\sqrt{2}}\left(\begin{array}{c}
\displaystyle\psi_{\mathbf{K}_{1}}(\mathbf{r})+\psi_{\mathbf{K}_{2}}(\mathbf{r})\\
\displaystyle\psi_{\mathbf{K}_{1}}(\mathbf{r})-\psi_{\mathbf{K}_{2}}(\mathbf{r})
\end{array}\right).
\label{eq:solution3}\end{equation}

Next we consider the case of $\gamma>1$, where the degeneracy of the SU(2) 
invariant condensate wavefunctions is lifted. 
In this case, within the convention of Eq.~(\ref{eq:bloch_expansion}), the 
solution of Eq.~(\ref{eq:solution3}) is selected, whose density, phase and 
spin distributions are plotted in Fig.~\ref{fig:density_phase_spin} ($c$) 
and ($d$). 
We see that bosons of different species occupy mostly the shallow sites 
in a checkerboard pattern with staggered spin density distribution. 
The condensate wavefunction in each species becomes real-valued with 
square-shaped nodal lines along with the period-doubled density profile, 
and we call this configuration the \textit{checkerboard} state.
In the single-species case \cite{Xu2013,cai2011}, the real non-Bloch states 
$\psi_{\mathbf{K}_{1}}(\mathbf{r})\pm\psi_{\mathbf{K}_{2}}(\mathbf{r})$ are always 
more energetic than the complex non-Bloch states
$\psi_{\mathbf{K}_{1}}(\mathbf{r})\pm i\psi_{\mathbf{K}_{2}}(\mathbf{r})$ and 
the real Bloch states $\psi_{\mathbf{K}_{1}}$ and $\psi_{\mathbf{K}_{2}}$, 
because the density distributions of the real non-Bloch states are less 
uniform than those of the latter ones. 
However, the conclusion is opposite in the two-species case: both species 
exhibit strong constructive and destructive interferences between 
$\psi_{\mathbf{K}_{1}}$ and $\psi_{\mathbf{K}_{2}}$ alternatively in adjacent 
shallow sites, and their real-space density distributions avoid 
each other and exhibit the checkerboard pattern. 
Consequently, the dominant interspecies interaction is greatly suppressed 
and the checkerboard state turns out to be the least energetic.

\begin{figure}[hbtp]\begin{center}
\includegraphics[width=0.9\linewidth]{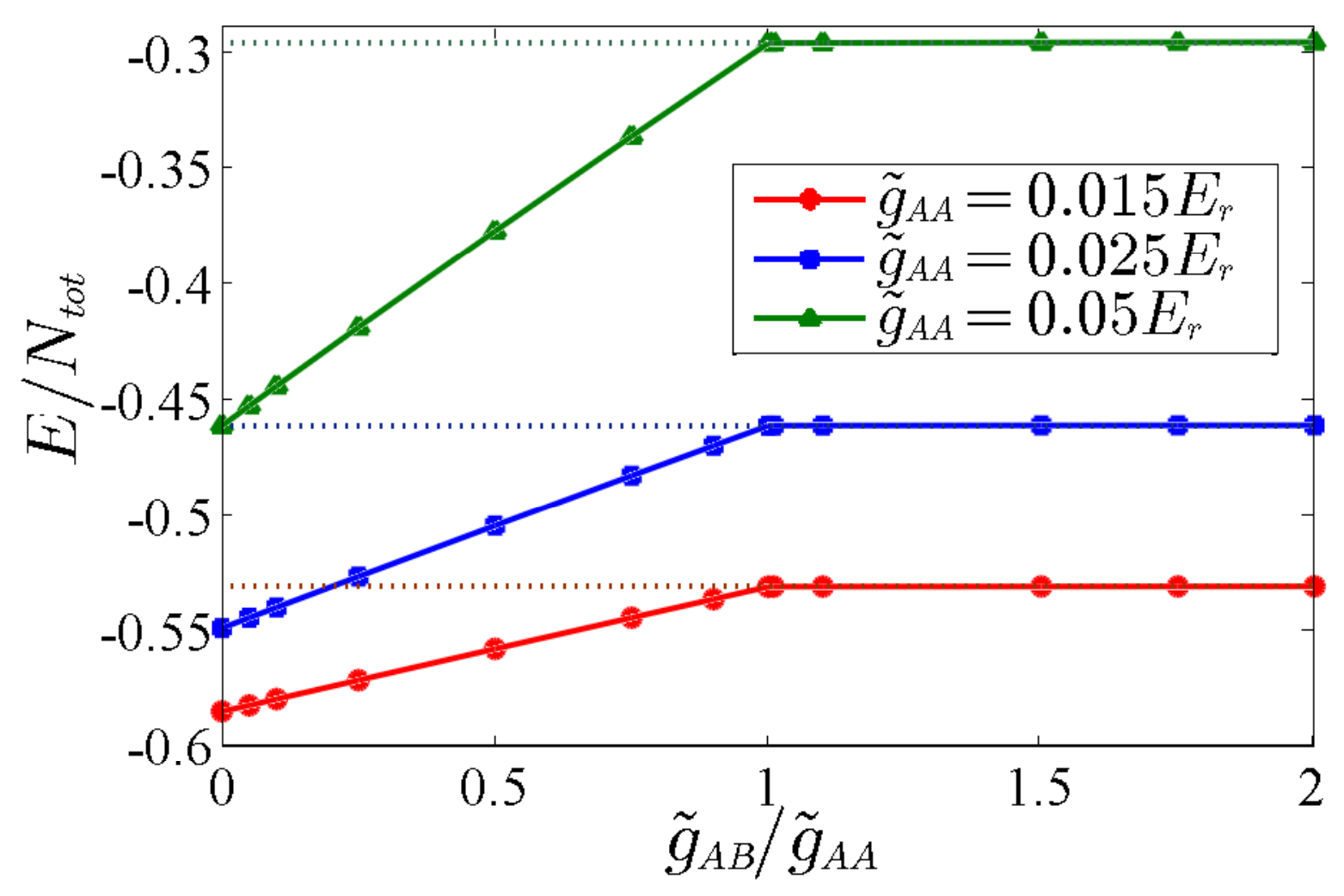}
\caption{$E/N_{tot}$ v.s. $\tilde{g}_{AB}/\tilde{g}_{AA}$ in the symmetric 
lattice with $\alpha=\alpha_0$.
Red dot, blue triangle, and green square are for different values of
$\tilde{g}_{AA}=0.015 E_r$, ~$0.025 E_r$, and $0.05 E_r$, respectively.
The dashed horizontal line represents the energy for $\gamma>1$ without i
ncluding domain walls, {\it i.e.}, solving the GP equation assuming 
fully polarization. The parameter values are the same as those in
Fig.~\ref{fig:density_phase_spin}.
}
\label{fig:energy}
\end{center}
\end{figure}

In the strongly repulsive regime ($\gamma>1$), however, it is possible that 
the system develops isolated "ferromagnetic" single-species domains. 
The case of spatial separation has been discussed for the bosonic mixture 
in the $s$-orbital bands of optical lattices in the same interaction 
regime~\cite{Altman2003,Mishra2007,Zhan2014,Lingua2015}.
When this scenario occurs in $p$-orbital bands, isolated domains of either 
species can choose themselves in whichever of the complex states, 
$\psi_{\mathbf{K}_{1}}(\mathbf{r})\pm i\psi_{\mathbf{K}_{2}}(\mathbf{r})$. 
We call such a configuration the spatially phase-separated spin-polarized state. 
Seemingly, this state could have an energy lower than that of the checkerboard 
state because of the vanishing interspecies interaction. 
In Fig.~\ref{fig:energy}, we plot $E/N_{tol}$ of full spin-polarized state 
with the complex condensate 
$\psi_{\mathbf{K}_1}(\mathbf{r})+i\psi_{\mathbf{K}_2}(\mathbf{r})$, or, 
its TR breaking counterpart.
Simple numerical test shows that the energy per particle of the checkerboard 
state is very close to that of the fully spin-polarized state.
When the initial state is prepared with $n_A=n_B$, the fully spin-polarized 
state becomes phase-separated spin polarization accompanied with the 
formation of inhomogeneous ferromagnetic domains, which cost the domain energy.
In spite of that, the checkerboard state of Eq.~(\ref{eq:solution3}) is still 
the prevailing UBEC state in this regime. 
Another issue is the time scale: Starting from the unpolarized initial state, 
forming ferromagnetic domains is a process of phase separation with a large 
scale arrangement of real space boson configurations.
It is much longer than the time scale of the formation of the checkerboard state
which only needs local phase adjustment of boson configurations.

\begin{figure}[tbp]\begin{center}
\includegraphics[width=1\linewidth]{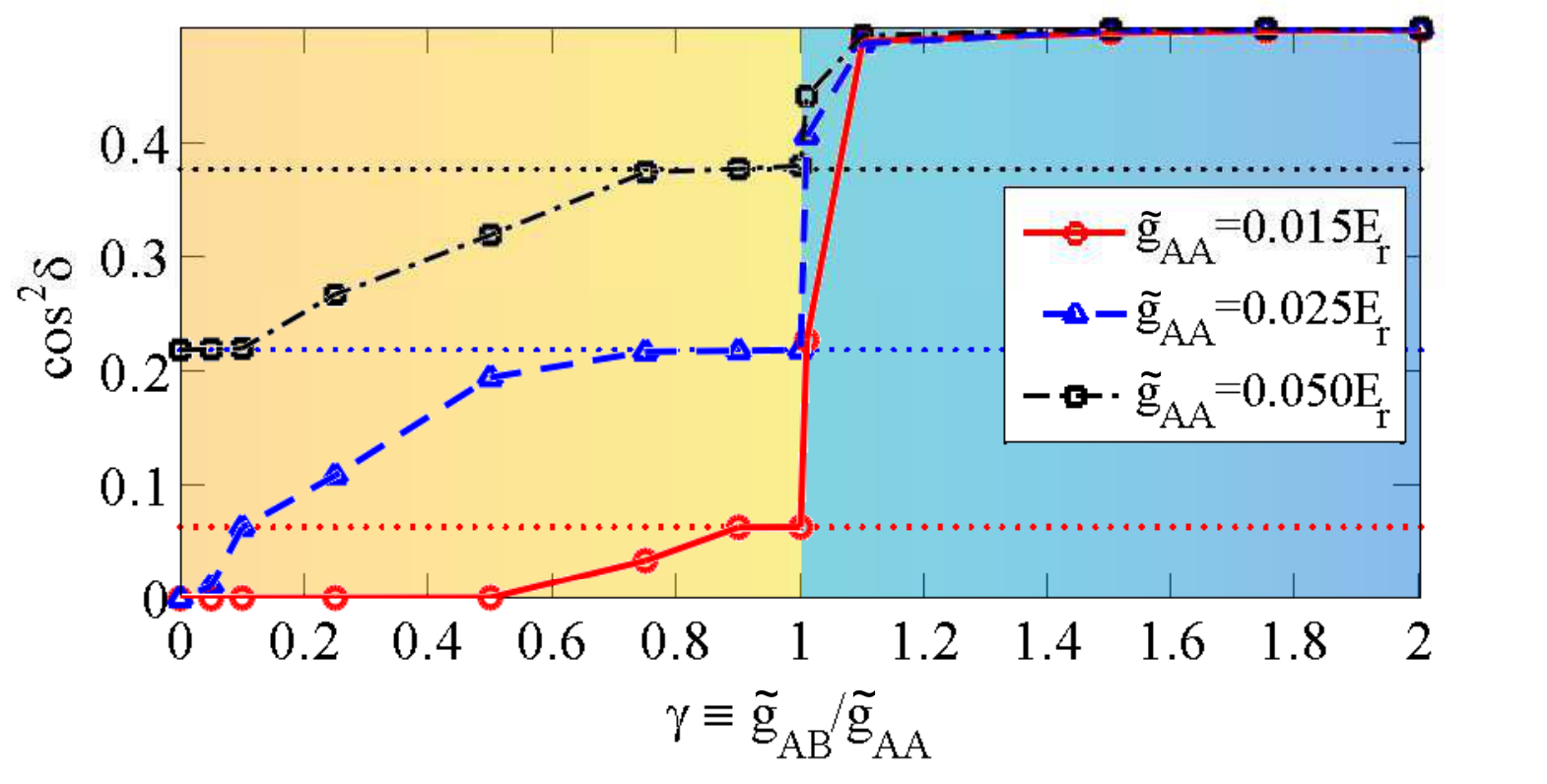}
\caption{The condensate fraction of $\psi_{\mathbf{K}_{1}}$, $\cos^{2}\delta$, 
as a function of $\gamma=\tilde{g}_{AB}/\tilde{g}_{AA}$.
The red dot, blue triangle and green square indicate 
$\tilde{g}_{AA}/E_{r}=0.015$, $0.025$ and $0.05$ respectively, 
and the dotted lines depict the fraction of spin-polarized state 
with the same total particle numbers for each interaction strength. 
The parameter values of the optical lattice here are the same 
as for Fig. \ref{fig:density_phase_spin} except for $\alpha=\pi/5>\alpha_{0}$.
}
\label{fig:assymmetry}
\end{center}
\end{figure}

\subsection{The asymmetric lattice}\label{sec:IVb}
\begin{figure}[tbp]\begin{center}
\includegraphics[width=1\linewidth]{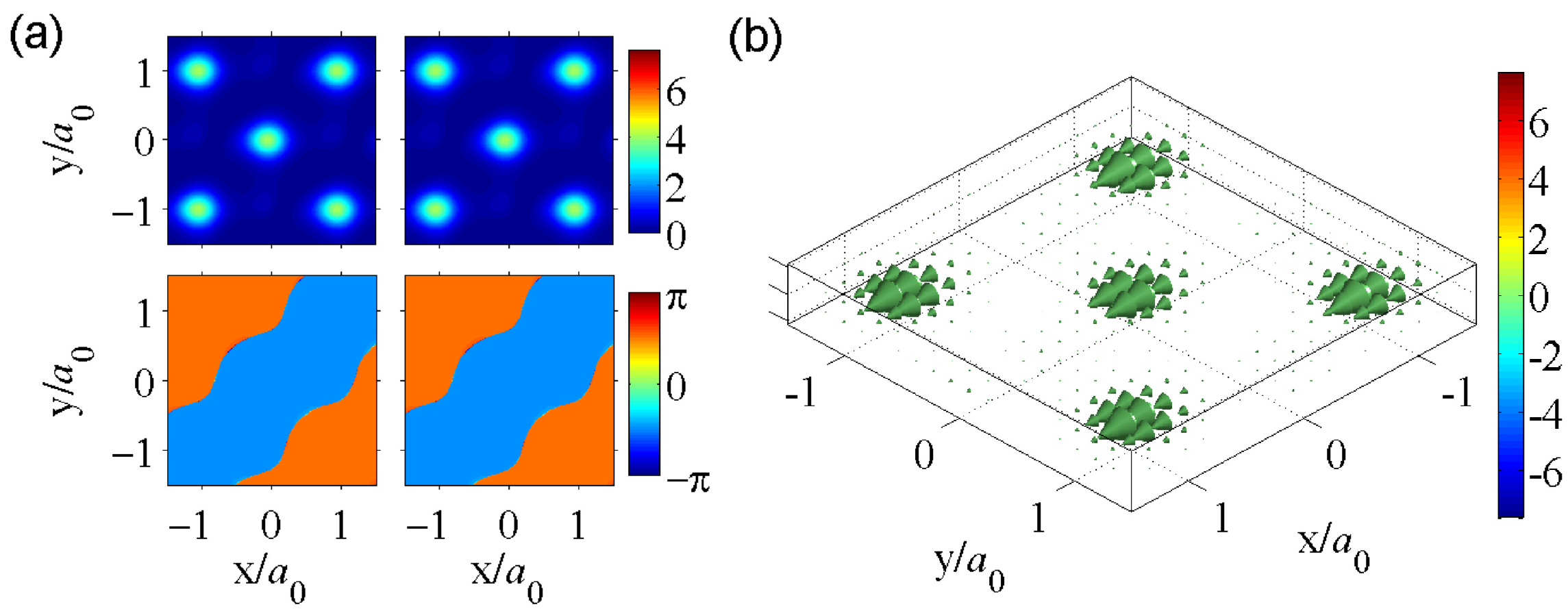}
\caption{(a) Density, phase and (b) spin texture of $\tilde{g}_{AA}=\tilde{g}_{BB} =0.015 E_r$, and $\gamma=0.1$. The parameter values of the optical lattice here are the same as for Fig.~1 in the main text except $\alpha=\pi/5$.}
\label{fig:real_small_int}
\end{center}
\end{figure}

Next we consider the interplay between lattice asymmetry and interactions. 
The lattice asymmetry breaks the degeneracy between the single particle 
states $\psi_{\mathbf{K}_{1}}$ and $\psi_{\mathbf{K}_{2}}$. 
Without loss of generality, we choose $\alpha>\alpha_{0}$, which sets the energy 
of $\psi_{\mathbf{K}_2}$ slightly lower than that of $\psi_{\mathbf{K}_1}$, such 
that the calculated condensate wavefunctions satisfy 
$\delta_{A}=\delta_{B}=\delta\neq\pi/4$. 
In Fig.~\ref{fig:assymmetry}, the condensate fraction of $\psi_{\mathbf{K}_{1}}$, 
$\cos^{2}\delta$, is plotted as a function of $\gamma$ at various values 
of $\tilde{g}_{AA}$. 
We find that the lattice asymmetry effect is more prominent for weak 
interactions. 
At $\tilde{g}_{AA}=0.015E_{r}$, the condensate fraction of 
$\psi_{\mathbf{K}_{1}}(\mathbf{r})$ vanishes when $\gamma<0.5$. 
The corresponding density, phase, and spin density distributions are depicted 
in Fig.~\ref{fig:real_small_int}~($a$) and ($b$). 
This is a real Bloch-type UBEC with a stripe-like configuration and an 
in-plane spin orientation. 
With increasing $\gamma$, $\psi_{\mathbf{K}_{1}}$ and $\psi_{\mathbf{K}_{2}}$ 
superpose in a complex way with $\phi_{A}=\phi_{B}=\pm\pi/2$ or 
$\phi_{A}=-\phi_{B}=\pm\pi/2$, but $\cos^{2}\delta$ remains small 
even at $\gamma=1$. We note that, only when $\gamma>1$, does 
the condensate quickly evolve to the checkerboard state. 
As $\tilde{g}_{AA}$ increases, the complex non-Bloch condensates become 
more and more prominent, as shown in Fig.~\ref{fig:assymmetry}.

Hitherto, we have concluded that the two-species $p$-orbital condensation can 
manifest itself in different forms of non-Bloch condensation: 
the complex states~(I) and (II), the real checkerboard state in 
Eq.~(\ref{eq:solution3}) and the spatially phase-separated spin-polarized state. 
Since all these states are linear combination of 
$\psi_{\mathbf{K}_{1}}(\mathbf{r})$ and $\psi_{\mathbf{K}_{2}}(\mathbf{r}),$ 
it is expected that four Bragg maxima would develop around the quasi-momenta, 
$\pm\mathbf{K}_{1,2}$ in the TOF spectra \cite{Liu2006,cai2011,wirth2011}. 
Given the condensate fractions $\psi_{\mathbf{K}_{1}}$ of Fig.~\ref{fig:assymmetry}, 
the states (I) and (II) as well as the spatially phase-separated complex 
spin-polarized state show that the relative intensities of these two pairs 
of peaks are dependent on the lattice asymmetry. 
However, when $\gamma>1$, the condensate fractions of $\psi_{\mathbf{K}_{1}}$ 
and $\psi_{\mathbf{K}_{2}}$ for the real checkerboard state quickly become 
nearly equally populated and thus the Bragg peaks of the TOF spectra have 
almost equal intensities, irrespective of the lattice asymmetry. 
This experimental observation could directly exclude the phase-separated 
spin-polarized state and provide supporting evidence for the phase 
transition from the complex UBECs towards the real-valued UBEC driven
by the interspecies interaction.

\section{Experimental scheme for phase measurement}\label{sec:V}

The two-species UBEC can be realized and observed by state-of-art experimental 
techniques~\cite{wirth2011,olschlager2011,olschlager2012,olschlager2013,Kock2015}.
Utilizing two different hyperfine spin states of an atom (labeled as the $A$- 
and $B$-species) \cite{myatt1997,Hall1998}, one first creates a condensate 
of sole species in the superposition of $\psi_{\mathbf{K}_{1}}(\mathbf{r})$ and 
$\psi_{\mathbf{K}_{2}}(\mathbf{r})$ which are the degenerate lowest-energy states 
in the $p$-orbital band. 
A $\pi/2$-Raman pulse is applied to convert half of the already condensed 
atoms into the other species. 
After tuning the interspecies atomic interaction with Feshbach resonance, the 
system is held for some time to let it relax to the intended non-Bloch 
$p$-orbital states, 
$\Psi_{A,B}=\left(\psi_{\mathbf{K}_{1}}+e^{i\phi_{A,B}}\psi_{\mathbf{K}_{2}}\right)$,
whose phase information can be inferred by matter-wave interferometry
as explained below.

After the preparation of the two-speicies condensate,
the atoms are then released from optical lattices and subsequently 
experience a Stern-Gerlach splitting during the ballistic expansion. 
Precisely, by applying a pulsed magnetic field gradient, the atoms
are accelerated by a spin-dependent force \cite{Machluf2013}, 
$\mathbf{F}_{\beta}\propto m_{\beta}|B|\hat{z}$ ($m_{\beta}$ is the 
projection of spin), and thus the two-species UBEC breaks into 
spatially separated parts along $z$ direction. 
A second $\pi/2$ Raman pulse is then applied to mix states of 
different momenta, leading to
\begin{equation}
\begin{array}{l}
\left(\begin{array}{c}
\tilde{\Psi}_{A}\\
\tilde{\Psi}_{B}
\end{array}\right)\propto\left(\psi_{\mathbf{K}_{1}}+e^{i\phi_{A}}\psi_{\mathbf{K}_{2}}\right)\left(\begin{array}{c}
1\\
ie^{i\Phi}
\end{array}\right)\otimes\left\vert \mathbf{p}_{A}\right\rangle \\
\\
\qquad\qquad+\left(\psi_{\mathbf{K}_{1}}+e^{i\phi_{B}}\psi_{\mathbf{K}_{2}}\right)\left(\begin{array}{c}
ie^{-i\Phi}\\
1
\end{array}\right)\otimes\left\vert \mathbf{p}_{B}\right\rangle
\end{array}
\label{expt_wf}
\end{equation}
where $\Phi$ accounts for the accumulated phases for the dynamical effects 
involved, and $\mathbf{\mathit{\mathbf{p}}}_{A,B}$ denote the momenta 
acquired by atoms after the Stern-Gerlach splitting. 
At this stage, the motion of each species is described by a wavepacket 
consisting of a superposition of two non-Bloch states with different 
quasi-momenta which interfere with each other along $z$ direction during 
the TOF \cite{Machluf2013,Kock2015}. 
The phase difference $\Delta\phi_{AB}=\phi_{A}-\phi_{B}$ can be inferred 
from the interference patterns imaged along vertical and horizontal 
directions for each species, as demonstrated in~\cite{Kock2015}. 
It can be shown from the Eq. (\ref{expt_wf}), that among the Bragg maxima, 
the $\mathbf{K}_{1}$ and $\mathbf{K}_{2}$ columns possess the same interference 
pattern, except the positions of fringes in the two columns are shifted by 
a phase angle, $|\Delta\phi_{AB}|$. 
By comparing the positions of fringes in the Bragg peaks, one can expect, 
when $\gamma<1$, $|\Delta\phi_{AB}|=0$ for state (I) and 
$|\Delta\phi_{AB}|=\pi$ for state (II).
Our scheme provides a feasible way for phase measurement in the current 
system.

\section{Summary and Discussions}\label{sec:VI}
In summary, we have studied the two-species $p$-orbital BECs in the 
experimentally accessible regime by a new imaginary-time propagation 
method for coupled GPEs, 
which can be applied to solve UBECs in higher bands. 
The competition between inter- and intraspecies interactions drives the 
transition from two non-equivalent complex-valued states, possessing 
respectively broken and unbroken TR symmetry, to a real-valued checkerboard 
state with a staggered spin density structure. 
We have also proposed experimental schemes to study the UBECs of the mixture. 
The current study paves the way for approaching the least explored $p$-orbital 
physics of multi-species bosonic systems. 
Our theory can be also generalized to study  the superfluity and magnetism 
of spinful $p$-orbital condensation in the presence of spin-dependent 
optical lattices or exotic spin-exchange interactions.

We have used the GPE method throughout this article, whose applicability
is justified in the limit of weak inter-species interaction.
In this case, the two-species problem studied here is reduced to two 
weakly coupled single-species problems, for which previous works show 
that the GPE equation has captured the essential physics of the 
complex $p$-orbital condensates being the energy-minima. 
When the interspecies interaction becomes stronger, however, the 
entanglement between two-species would become important. 
In this case, indeed more exotic states beyond the GPE level is also possible. 
For example, the singlet paired boson condensation, whose spatial 
pair-wavefunctions are antisymmetrized, and thus reduces 
the inter-species repulsion. 
This state is highly entangled and beyond the GPE equation level. 
Nevertheless, the mean-field GPE equation is still a natural beginning 
point on this challenging problem. 
The checkerboard state already investigated in this article remains 
a potential competing state,
then both species avoid each other in their real-space density 
distributions characterized by a staggered spin-density structure, which 
also greatly reduces the interspecies repulsion. 
We would leave a detailed study on novel states beyond the mean-field GPE 
level and their competitions with the single-boson condensate for a 
future publication.

\textit{Acknowledgments} 
J.-S. Y. is supported by the Ministry of Science and Technology, Taiwan 
(Grant No. MOST 102-2917-I-007-032). I.-K. L. and S.-C. G. are supported by 
the Ministry of Science and Technology, Taiwan (Grant No. MOST 103-2112- 
M-018- 002-MY3). S.-C.G. is also supported by the National Center for 
Theoretical Sciences, Taiwan. C. W. is supported by the NSF DMR-1410375 
and AFOSR FA9550-14-1-0168. C. W. acknowledges the support from 
President's Research Catalyst Award No. CA-15-327861 from the University 
of California Office of the President. 
We also acknowledge M.-S. Chang for his comments on the proposed 
experimental scheme.


\begin{thebibliography}{10}
\bibitem{species_dependent}
Here, we have assumed that both species share the same energy spectrum of
  $p$-band and hence $\psi_{\mathbf{K}_{1,2}}$ due to a totally symmetric
  setting. However, such a degeneracy is lifted when imbalance in the atomic
  mass or population occurs, that $\psi_{\mathbf{K}_{1,2}}$ turn out to be
  species-dependent.
\bibitem{su2rotation}
The SU(2) rotations are defined as
  $\hat{R}=\exp(-i\phi\hat{\boldsymbol{\sigma}}\cdot{\mathbf{n}}/2)$ with $\phi$ the
  rotation angle around ${\mathbf{n}}$, and $\hat{\boldsymbol{\sigma}}$ the Pauli vector.
\end{thebibliography}

\end{document}